\documentclass[12pt,a4paper,epsf]{article}
\usepackage{graphics}
\usepackage{amssymb,amsmath}
\usepackage[dvips]{lscape,graphicx}
\usepackage{cite}
\usepackage{longtable}
\usepackage{color}
\usepackage{slashed}
\usepackage{authblk}
\newcommand{\bc}{\begin{center}}
\newcommand{\ec}{\end{center}}
\newcommand{\bd}{\begin{displaymath}}
\newcommand{\ed}{\end{displaymath}}
\newcommand{\be}{\begin{equation}}
\newcommand{\ee}{\end{equation}}
\newcommand{\ba}{\begin{array}}
\newcommand{\ea}{\end{array}}
\newcommand{\bea}{\begin{eqnarray}}
\newcommand{\eea}{\end{eqnarray}}
\newcommand{\bt}{\begin{tabular}}
\newcommand{\et}{\end{tabular}}

\newcommand{\bp}{\begin{picture}}
\newcommand{\ep}{\end{picture}}
\newcommand{\bfi}{\begin{figure}}
\newcommand{\efi}{\end{figure}}

\def\fun#1#2{\lower3.6pt\vbox{\baselineskip0pt\lineskip.9pt
\ialign{$\mathsurround=0pt#1\hfil##\hfil$\crcr#2\crcr\sim\crcr}}}

\begin{document}



\vspace{1cm}

\title{\LARGE \bf Magnetic Moment of
Constituent Fermions in Strongly Interacting Matter}
\author{
H.B.~Nielsen
\footnote{hbech@nbi.dk}}
\affil{Niels Bohr Institute, Copenhagen, Denmark.}
\author{Vikram Soni
\footnote{v.soni@airtelmail.in}}
\affil{Centre for Theoretical Physics, Jamia Millia Islamia, New Delhi, India.}

\maketitle

\thispagestyle{empty}

\begin{abstract}
We investigate  the magnetic 
moment 
operator for constituent fermion
masses for chirally symmetric theories. 
Constituent fermion masses are generated 
through 
a yukawa interaction
 of the fermion with a scalar (or /and 
psuedoscalar) field via the vacuum 
expectation value (VEV) of the scalar 
(or and psuedoscalar) field.

 We especially consider
 the high baryon density $\pi_0$ 
condensed phase, in which chiral symmetry 
is  spontaneously
 broken,  with space varying  expectation 
values of the $ \sigma$ and  $\pi_0$ 
fields.
 This phase has a spin polarized fermi 
sea as the ground state. We show that 
there is indeed
generated a macroscopic magnetization in 
this phase, contrary to what one would 
have found, if one just used
a primitive phenomenological magnetic 
moment formula for explicit/ current  
fermion masses.

Furthermore, this analysis reveals  that 
the magnetization of this state goes up 
as the VEV, that determines the 'mass',
comes down with increasing baryon 
density. The consequent high magnetic field 
that is generated will destabilize this state
 at a threshold density. This is  
important in the context of neutron 
stars, as such a high density state may 
be responsible for very high magnetic 
fields in the dense core of neutron stars.
This could potentially be the origin of magnetars - 
the stars with the largest magnetic 
fields in the universe.
\end{abstract}
\clearpage\newpage

\section{Introduction}


Constituent masses are generated via 
yukawa interactions of the fermion with a 
scalar (or /and psuedoscalar) field via 
the vacuum expectation value (VEV) of the 
scalar (or and psuedoscalar) field. 
Constituent fermions arise in a variety of
models from the Standard model of the
electroweak interactions via the Higgs VEV
or in the strong interactions(in the
 chiral symmetry limit) where
the nucleon / quark masses are generated
by the VEV of the sigma-field  $<\sigma>$.
Constituent masses are different from 
explicit or current masses. The obvious 
difference is that under a
chiral symmetry transformation the exact 
chiral symmetric Hamiltonian remains 
invariant (constituent masses) whereas
any current mass in the Hamiltonian 
explicitly breaks the chiral symmetry.


 Also, there are physical situations in 
which  the VEV's of the scalar (or and 
psuedoscalar) field and therefore 
constituent masses can vary, like at 
finite density or temperature, which 
underlines the difference with current 
masses which 'do not' change with finite 
density or temperature. Indeed, there are 
situations where the VEV's can even 
become space varying, making the 
difference with current masses even more 
dramatic. In such a context the use of 
the usual  phenomenological formulae(e.g. for magnetic moment), 
that do not take account of this, can be misleading. 

One such case that we will 
consider is the case of a high baryon 
density ground state for strongly 
interacting nucleons, which has a  
$ \pi_0 $  condensate that is a 
(space varying) stationary 
wave\cite{dautry,baym77,kutschera+90,
dbvsplb}. In this case when we calculate 
the magnetic moment of the nucleons in 
the presence of a $ \pi_0 $  condensate using the naive formula,
we find a somewhat strange result: that 
for a spin polarised neutron ground 
state, the magnetic moment vanishes when 
averaged over space\footnote{R. F. Sawyer private communication}.  
However, when we compute the 
magnetic moment operator from first 
principles, from the chiral lagrangian, we find it has the right space 
dependence, 
which cancels out with the space 
dependence of the ground state to give a 
magnetic moment that is proportional to 
the total spin. Furthermore, this 
reveals  that the magnetization of this 
state goes up as the VEV, that determines 
the 'mass', comes down with increasing 
baryon density..

This is important in the context of 
neutron stars, as such a high density 
ground state may be responsible for very high 
magnetic fields in the dense core and  could be the origin of 
magnetars \cite{dbvsarxiv,hds}- the stars 
with the largest magnetic fields in the 
universe.

\section{ Current and Constituent  Masses}

\subsection{QED }
 Quantum Electrodynamics is a theory with an explicit
or current electron mass, $m_e$, which breaks chiral symmetry explicitly. 
The case of the electron magnetic moment
operator is explicitly worked out in the 
text of Sakurai (Advanced Quantum 
Mechanics) via the
Gordon decomposition 
(see (3-3 page 85)\cite{sak}).

\begin{equation}
 H_ {mag}= -\frac{e}{2m_e}\frac{1}{2} F_{\mu\nu}
\bar{\psi}\sigma_{\mu\nu} \psi. \label{f1}
\end{equation}
($\hbar = c =1$)


\subsection {The Chiral symmetric Gellmann Levy  sigma model}

In a chiral symmetric  theory, 
the mass of the
nucleon/quark comes from the VEV of
$<\sigma>$, for example  in the linear
$\sigma$-model of Gellman and Levy[Gellmann Levy 1960].
In the case of exact chiral symmetry we 
have the following lagrangian when we 
couple the Gellmann Levy sigma model to 
the electromagnetic field

\begin{equation}
  L =
 - \frac{1}{4} F_{\mu \nu} F_{\mu \nu}   \\
 - \sum {\overline{\psi}} 
\left( \slashed{D} + g_y(\sigma +  \\
i\gamma_5 \vec \tau \cdot\vec \pi)\right) 
\psi            \\
- \frac{1}{2} (\partial                                \\
\mu \sigma)^2 - \frac{1}{2} (\partial 
\mu \vec \pi)^2    \\
-  \frac{\lambda^2}{4}(\sigma^2 + 
\vec \pi^2 - (F)^2)^2 \\
\end{equation}

In the limiting case of a small explicit 
pion mass being set to zero, the usual (uniform in space) symmetry breaking 
that follows on the
minimization of the potentials above is,  
$ <\sigma> = F = f_\pi $,and  
$<\vec\pi> = 0$.

The masses of the scalar (PS) and 
fermions  are given by

\begin{equation}
\qquad  <\sigma>^2 = F^2
\end{equation}
where $F$ is the pion decay constant.
It follows that
\begin{equation}
\qquad m^2_{\sigma} = 2\lambda^2 F^2
\qquad  m = g <\sigma> = g F
\end{equation}

The quark or nucleon mass m, is a spontaneous  mass that is generated from 
the spatially uniform VEV. 
In this case the usual magnetic moment formula above (\ref{f1}) works.

On the other hand when the VEV's of $ <\sigma>$ and $<\vec\pi>$ depend 
on space  coordinates, the above expression for the magnetic moment will not 
work as we will find in the following section.

\section { The $ \pi_0 $ 
condensation : space dependent VEV's}

\subsection{The $ \pi_0 $ 
condensed ground state}

Here we shall consider another realization
of the expectation value of  $<\sigma>$ 
and $<\vec\pi>$ corresponding
to 
$\pi_0$ condensation. This 
phenomenon was first considered in the 
context
of nuclear matter \cite{dautry,baym77}. 
Such a phenomenon also occurs with our 
quark based chiral $\sigma$
model and was  considered at the mean 
field level by Kutschera, Broniowski and 
Kotlorz for the 2 flavour case 
\cite{kutschera+90} and by one of us for 
the 3 flavour case \cite{dbvsplb}. 
Working in the
chiral limit they found that the pion 
condensed
state has lower energy than the uniform 
symmetry breaking state
we have just considered 
for all density. 
This is expected, as the ansatz
for the pion condenced  phase is more general.

The expectation values now carry a 
particular space dependence
\begin{eqnarray}
      <\sigma> &=& F  
\cos{(\vec q. \vec r)} \\
      <\pi_3> & =&  F  
\sin{(\vec q. \vec r)} \\
      <\pi_1> &=& 0 \\
      <\pi_2> &=& 0
\end{eqnarray}

Note that the relation, $<\sigma^2> + <\vec \pi^2> = F^2$, is preserved under 
this pattern of symmetry breaking.
Also, when $|\vec q |$ goes to zero, we recover the usual space uniform phase 
above.

The Dirac Equation in this background is 
solved in \cite{dautry,kutschera+90} by 
the artifact of writing the wavefunction, 
$\psi $, in terms of a chirally rotated 
wave function, $\chi(k)$,
\begin{equation}
 \psi = \exp(- i ({\tau_3}/2){\gamma_5}
{\vec q}\cdot {\vec r})\cdot\chi(k) 
\end{equation}
  where, $\chi(k)$ are momentum 
eigenfunctions.

The Hamiltonian reduces to
\begin{equation}
     H \chi(k) = (\vec \alpha . 
\vec k - \frac{1}{2} \vec q . \vec \alpha 
\gamma_5 \tau_3 + \beta m) \chi (k) 
= E(k)\chi(k)
\end{equation}
where $m  = g\sqrt{<\sigma>^2 + 
<\vec \pi>^2} = g F $

The second term arises from the condensate and has been written in terms 
of the relativistic spin
operator, $\vec \alpha \gamma_5$ .  It is 
evident that if spin is parallel to 
$ \vec q$ and $ \tau_3 = +1$
(proton/up quark) this term is negative 
and if $\tau_3 = -1 $ (neutron/down 
quark) it is positive.
For spin antiparallel to $\vec q$ the 
signs of this term for $\tau_3 = +1$ and $-1$  are
reversed.

The spectrum for the hamiltonian is the 
quasi particle spectrum and
can be found to be \cite{dautry,kutschera+90}
\begin{eqnarray}
      E_{(-)}(k) &=& \sqrt{ m^2 + k^2 +
\frac{1}{4}q^2 -\sqrt{m^2q^2 +
                 (\vec q.\vec k )^2}} \\
 E_{(+)}(k) &=&  \sqrt{m^2 + k^2 + 
\frac{1}{4}q^2 + \sqrt{m^2q^2 +
                     (\vec q.\vec k)^2}}
\end{eqnarray}
The lower energy eigenvalue $ E_{(-)}$ has 
spin along $ \vec q$ for
$ \tau_3 =1$, or has spin  opposite to 
$\vec q$ for $\tau_3 =-1$.
The higher energy eigenvalue
$E_{(+)}$ has spin along $\vec q$ for 
$\tau_3 = -1$, or has spin
opposite to $\vec q $ for $ \tau_3 = +1$.

 The ground state is constructed by 
occupying all 
the lower energy states $ E_{(-)}$ as 
the gap between the lower and higher 
states is large. In  this  background the 
fermi sea is no longer degenerate in spin
but gets polarized into the  states above.



\subsection{The naive magnetic 
moment calculation in the 
$\pi_0$ condensed state}
Let the condensate wave vector 
$ \vec q $, define the $\hat{z}$ axis. For 
neutrons $\tau_3 = -1$. Since the spins 
are aligned opposite to  $ \vec q $, we 
expect that there should be a net 
magnetic moment opposite to the 
$ \vec q $ axis . Let us look at the 
source of the magnetic field. from the 
`usual' Gordon decomposition of the electro magnetic  
current\cite{sak}.
  \begin{equation}
      H_{mag} =
 \frac{e}{2m}  F_{\mu\nu}\cdot 
{\overline{\psi}} \left(\sigma_{\mu,\nu}
\tau_{3}\right) \psi            \\
    \end{equation}

where, $ m = 
g \sqrt{<\sigma>^2 + <\vec \pi>^2} $.

 Substituting for  $\psi = \exp(- i ({\tau_3}/2)
{\gamma_5}{\vec q}\cdot {\vec r})\cdot
\chi(k) $,
  where, $\chi(k)$ are momentum 
eigenfunctions, we find

  \begin{equation}\label{f14}
       H_{mag}  =
  \frac{e}{2m} F_{\mu\nu}\cdot 
{\overline{\chi(k)}} \sigma_{\mu,\nu}\tau_{3}
 \exp(- i \tau_{3}{\gamma_5}{\vec q}
\cdot {\vec r}) \chi(k)
        \\
\end{equation}

 The ground state is constructed by filling the spin polarized  
  $ E_ {-} (k)$  states up to the fermi momentum.  
The total spin of the ground state is the sum over the spin polarized 
Fermi-sea.  
Do the magnetic moments also add up like the spins?

If we simply used   the normal(explicit current mass) formula (\ref{f14}) we 
would  
get a space average over the above
$\sin{\vec{q}\cdot\vec{x}}$ and $\cos{\vec{q}\cdot\vec{x}}$ terms that 
goes to zero.

\subsection{ The full chiral  magnetic moment calculation in the $\pi_0$ 
condensed state}

Using the notation of Sakurai\cite{sak}, 
\hbox{$ \gamma_\mu\partial_\mu  =  
\gamma_i\partial_i + \gamma_4\partial_4
 = \gamma_i\partial_i + 
(i\gamma_0)(-i\partial_0) $}
we may write the following chiral Dirac 
equations for the the $\pi_0$ condensed 
state

\begin{align}
\big(\gamma_\mu\partial_\mu  
-ie\gamma_\mu A_\mu + g (\sigma +
i\gamma_5 \vec \tau \cdot\vec \pi)\big) 
\psi   &= 0\\
\big(\gamma_\mu\partial_\mu  
-ie\gamma_\mu A_\mu + 
m \exp(i {\tau_3}{\gamma_5}{\vec q}\cdot 
{\vec r})\big)
\psi  &= 0
\end{align}

Now we can invert this equation to get
\begin{equation}
\psi   =  - \frac{1}{m}\exp(- i \tau_{3}{\gamma_5}{\vec q}\cdot {\vec r})  
({\gamma_\mu}{\partial_\mu}  -ie{\gamma_\mu} A_\mu )\psi \label{f17}
\end{equation}

Similarly, by complex conjugation with a little algebra  we find

\begin{equation}
\overline{\psi}   =   \frac{1}{m} ({\partial_\mu}{\overline{\psi}}{\gamma_\mu }
  + ie{\overline{\psi}}{\gamma_\mu} A_\mu ))\exp(- i \tau_{3}{\gamma_5}
{\vec q}\cdot {\vec r})\label{f18}
\end{equation}

Following the 'Sakurai' trick \cite{sak}, we use the ploy of dividing the 
electro magnetic  current operator into two identical parts

 \begin{equation}
 j_{\mu}  = \frac{e}{2}[\overline{\psi}{\gamma_\mu}\psi   +   
\overline{\psi}{\gamma_\mu}\psi ]\\ \label{f19}
 \end{equation}

 We then use the chiral Dirac equation(\ref{f17}) above for 
$\overline{\psi}$, and insert it in the first term  and (\ref{f18}) for , $\psi$, and insert it in 
the second term.

 After doing the Gordon decomposition, we get the following expression 
 for the electromagnetic current operator (if we leave out all the terms 
that depend on the vector potential, $ A_{\mu} $), 

 \begin{align}
  j_{\mu}   &=  \frac{e}{2m}   {\bf {\huge  (}}   [(\partial_{ \nu} 
{\overline{\psi}})  
\exp( i \tau_{3}{\gamma_5}{\vec q}\cdot {\vec r}) \sigma_{\mu,\nu}\tau_{3}
\psi +  {\overline{\psi}}  \exp( i \tau_{3}{\gamma_5}{\vec q}\cdot {\vec r}) 
\sigma_{\mu,\nu}\tau_{3}(\partial_{ \nu} \psi)] \nonumber\\
  &\quad +i [(\partial_{ \mu} {\overline{\psi}})  \exp( i \tau_{3}{\gamma_5}
{\vec q}\cdot {\vec r}) \tau_{3}\psi - {\overline{\psi}}  
\exp( i \tau_{3}{\gamma_5}
{\vec q}\cdot {\vec r}) \tau_{3}(\partial_{ \mu} \psi) ] {\bf {\huge  )}}
 \end{align}

 Notice the presence of the phase factor in the current operator. The 
expression reduces to the usual operator,  if 
we put  $ q  = 0 $ ,for then the phase factor drops out $ ( = 1) $. The first 
part has the Lorentz structure of the magnetic moment operator and  the second 
term reduces to a current density in the Schroedinger theory\cite{sak}. For our present purpose we will concentrate on the first term which is the magnetic moment type term.

 \begin{align}
 H_{mag}  = - j_{\mu}^{mag} A_{\mu}  &= - \frac{e}{2m}  A_{\mu} 
[\partial_{ \nu} ( {\overline{\psi}}  \exp( i \tau_{3}{\gamma_5}{\vec q}\cdot {\vec r}) \sigma_{\mu,\nu}\tau_{3}\psi )\nonumber \\
 &\quad - \overline{\psi} (i {q_{\nu}}{\gamma_5} \tau_{3})\exp( i \tau_{3}{\gamma_5}{\vec q}\cdot {\vec r}) \sigma_{\mu,\nu}\tau_{3}\psi ]
 \end{align}

  By partial integration \textsl{of the first term} above, we can move the derivative to act on the vector potential , $ A_\mu $, and write it in the usual form of the magnetic moment term. However, now there is an extra term, the second term, \textsl{which needs a physical interpretation}.

\hspace{3mm}

\textbf{The first term}

 The first term is the regular magnetic moment term and is
  \begin{equation}
   H_{1mag} = \frac{e}{4m}  F_{\mu,\nu} [\overline{\psi}  \exp( i \tau_{3}{\gamma_5}{\vec q}\cdot {\vec r})\sigma_{\mu,\nu}\tau_{3}\psi ]
 \end{equation}

 On substituting for $\psi = \exp(- i ({\tau_3}/2){\gamma_5}{\vec q}\cdot {\vec r})\cdot\chi(k) $
 we find that the two phase factors, from the magnetic moment operator and the wavefunction, nicely cancel  to give

 \begin{equation}
  \frac{e}{2m}  F_{\mu\nu}\cdot {\overline{\chi(k)}} \sigma_{\mu,\nu} \chi(k)
\end{equation}

This magnetic moment is \textsl{not} 
space 
dependent and does not average to zero 
and thus restores  the proportionality between the total 
spin and  total magnetic moment for the ground state.

\hspace{3mm}
\textbf{The second term}

\begin{align}
   H_{2mag} = - ( \frac{e}{2m} ) A_{\mu} 
[ - \overline{\psi} (i {q_{\nu}}{\gamma_5} 
\tau_{3})\exp( i \tau_{3}{\gamma_5}{\vec q}
\cdot {\vec r}) \sigma_{\mu,\nu}
\tau_{3}\psi ]
 \end{align}

By using the field equations above 
for $\psi$ and $\overline{\psi}$  and 
the 'Sakurai trick' \cite{sak} the 
second term can be recast to yield many 
terms. 

We drop the terms that are not relevant for an effective magnetic 
moment term for reasons of economy. We also drop terms 
which are quadratic or higher order in  'q' and /or $A_{\mu}$.
We are then left with only a single term,
which has the form of a magnetic moment operator.
Below we write this term in terms of $\chi$:

 \begin{align}( i\frac{e}{8m^2} ) 
F_{\mu\lambda}\cdot {\overline{\chi(k)}}
{q_{\nu}}{\gamma_5} \tau_{3} 
\sigma_{\mu,\nu} {\gamma_{\lambda}}\chi(k)
 \end{align}

This term is a somewhat different 
Dirac bilinear that interacts with the 
magnetic field.We note in passing  that 
the electromagnetic interaction is not 
invariant under chiral transformations. 

\section{The magnetic field in 
the $\pi_0$ condensation phase 
transition as a function of 
baryon density}

The interesting question that this poses 
in the context of the  $\pi_0$ 
condensation phase transition is  how does 
the chiral mass term behave as a function 
of baryon density? It can be seen 
from fig 4 \cite{kutschera+90} that as 
the density goes up the value of the 
chiral mass, $m  = g\sqrt{<\sigma>^2 + 
<\vec \pi>^2}  $ , 
decreases monotonically. As this mass 
appears in the denominator of the 
expression for the magnetic moment, the 
magnetic moment keeps going up as the 
density increases. This implies that 
the $ \pi_0$ condensed state in which 
spins and magnetic moments are aligned  
would have higher magnetic moments and 
therefore higher magnetic fields as the 
density goes up.

Will the effect keep increasing as 
density goes up or is there a cap on the 
magnetic field?
As baryon density is increased the 
chiral mass parameter goes down, lowering 
the condensate energy but simultaneously 
increasing the magnetic energy.
An 
upper bound to the magnitude of the 
allowed magnetic field can be obtained 
by equating the condensation energy 
density to the magnetic energy density

\begin{equation}
  (\lambda^2/4) \cdot (\frac{m}{g})^4   
\simeq  B^2/(8\pi)
\end{equation}

Kutschera et al  
\cite{kutschera+90}  use a sigma 
particle mass of $1200 Mev$ corresponding to  
$\lambda^2 \simeq 75$. For a more 
realistic sigma particle mass of 
$600 Mev$ we have $\lambda^2 \simeq 20 $. 
From the figure 4 in \cite{kutschera+90}
the value of the chiral mass parameter 
at a baryon density of $ \simeq 1/{F^3}$ is about half its value at 0 
density ($m = g f_{\pi}$). This yields an upper bound for the magnetic field 
that 
the condensate can sustain, $\simeq  10^{17}$ gauss, which is close to the 
field that we found $\simeq  10^{16}$ gauss \cite{dbvsarxiv,hds}, as the 
core field in our magnetar model. The implication is that in  the $ \pi_0$ 
condensed state in  the core of the star we are 
close to upper critical field. Beyond this 
density/magnetic field the ground state is not stable and is likely to 
transform into a charged pion condensate or unpolarized fermi sea 
( which carry no magnetic moment) - an interesting inference.

\section{Acknowledgement}
We thank Ray Sawyer for pointing out this discrepancy with the  magnetic 
moment operator in a review of \cite{hds} 
VS acknowledges discussions with 
N. D. Haridass. One of us (H.B.N.) thanks
Indian Institute of Technology in Roorkee
and NIT Warangal Technical Institute for 
bringing him to India, Mitja Breskvar 
for bringing him to Bled Conference,
and the Niels Bohr Institute for an
 emeritus position.  
 We also thank the Niels Bohr 
Institute for 
supporting the visit   of 
the other author(V.S.).

\end{document}